\documentclass[final,5pt,twocolumn,amssymb,prd]{revtex4-1}
\usepackage{pgfplots}
\usepackage{latexsym,epsfig}
\usepackage{graphicx}
\usepackage{ulem}
\usepackage{dcolumn}
\usepackage{color}
\usepackage{amsthm,amsmath}

\begin{document}

\title{New Physics Constraints from Atomic Parity Violation in $^{133}$Cs}

\author{B. K. Sahoo$^a$\footnote{Email: bijaya@prl.res.in}, B. P. Das$^{b,c}$ and H. Spiesberger$^d$}
\affiliation{$^a$Atomic, Molecular and Optical Physics Division, Physical Research Laboratory, Navrangpura, Ahmedabad-380009, India\\
$^b$Department of Physics, School of Science, Tokyo Institute of Technology, 2-1-2-1-H86 Ookayama Meguro-ku, Tokyo 152-8550, Japan \\
$^c$Centre for Quantum Engineering Research and Education, TCG Centres for Research in Science and Technology, Sector V, Salt Lake, Kolkata 70091, 
India \\
$^d$PRISMA$^+$ Cluster of Excellence, Institut f\"{u}r Physik, Johannes Gutenberg-Universit\"{a}t, D-55099 Mainz, Germany}

\begin{abstract}
Our improved calculation of the nuclear spin-independent parity violating electric dipole transition amplitude ($E1_{PV}$) for $6s ~ ^2S_{1/2} - 
7s ~ ^2S_{1/2}$ in $^{133}$Cs in combination with the most  accurate (0.3\%) measurement of this quantity yields a new value for the nuclear 
weak charge $Q_W=-73.71(26)_{ex} (23)_{th}$ against the Standard Model (SM) prediction $Q_W^{\text{SM}}=-73.23(1)$. The advances in our 
calculation of $E1_{PV}$ have been achieved by using a variant of the perturbed relativistic coupled-cluster theory which treats the contributions 
of the core, valence and excited  states to $E1_{PV}$ on the same footing unlike the previous high precision calculations. Furthermore, this 
approach resolves the controversy regarding  the sign of the core correlation effects. We discuss the implications of the deviation of our result 
for $Q_W$ from the SM value by considering different  scenarios of new physics.
\end{abstract}

\date{Received date; Accepted date}

\maketitle

The neutral current weak interactions due to the exchange of a $Z_0$ boson between the electrons and the nucleus in an atom leads to parity violation \cite{Bouchiat}.
This phenomenon is referred to as atomic parity violation (APV). The nuclear spin-independent (NSI) APV has been measured to an accuracy of 0.35\% in the $6s ~ ^2S_{1/2} - 7s ~ 
^2S_{1/2}$ transition in $^{133}$Cs \cite{wood}. This is the most accurate APV measurement to date, but two recent proposals 
\cite{choi,anders} have the potential to surpass this accuracy. Thus, the stage is now clearly set to take the APV calculations in Cs to the 
next level. This indeed provides the motivation for our present work. The principal quantity of interest in the APV studies is the nuclear weak charge
(NWC), which is a linear combination of the NSI neutral current weak interaction coupling coefficients between electrons and up- and down-quarks in
an atom \cite{Bouchiat,bouchiat1}. The difference in the model independent value of NWC obtained from APV and that obtained from the Standard Model
(SM) could shed light on new physics beyond the SM (BSM). 

 Following the conventions using 4-fermion operators defined in the PDG \cite{PDG2020}, the parity-violating lepton-hadron interactions at low 
energies can be described by 
{\small{
\begin{eqnarray}
\mathcal{L}^{\mathrm{NP}} &=& \frac{G_{F}}{\sqrt{2}} \left[ \overline{e} \gamma^{\mu} \gamma^{5} e 
 \sum_q g_{AV}^{eq} \overline{q} \gamma_{\mu} q  + \overline{e} \gamma^{\mu} e \sum_q g_{VA}^{eq} \overline{q} \gamma_{\mu} \gamma^{5} q \right], \ \ \
\label{eq:4f-PDGnew}
\end{eqnarray}
}} 
where $G_F$ is the Fermi constant and the sum over $q$ includes the interaction of electrons with up ($u$), down 
($d$) and strange ($s$) quarks. Note that in earlier editions of the PDG~\cite{Beringer:1900zz} a different notation was used, writing $C_{1 q}$ 
instead of $g_{AV}^{eq}$ and $C_{2 q}$ instead of $g_{VA}^{eq}$. The coupling constants $g_{AV}^{eq}$ and $g_{VA}^{eq}$ are defined in the static 
limit and are universal. The temporal component of the quark currents determines the NSI weak interaction Hamiltonian which is 
used in the atomic calculations. Assuming that the nucleons can be treated non-relativistically and point-like, Eq.~(\ref{eq:4f-PDGnew}) allows us to 
define parity-violating couplings for protons and neutrons at vanishing momentum transfer, 
\begin{equation}
g_{AV}^{ep} = 2 g_{AV}^{eu} + g_{AV}^{ed} \,  
\quad \text{and} \quad 
g_{AV}^{en} = g_{AV}^{eu} + 2 g_{AV}^{ed} \, . 
\end{equation} 
These add up coherently across the nucleus giving rise to the NWC of the nucleus. With $Z$ protons and $N$ neutrons, the NWC at leading order 
is given by 
\begin{equation}
Q_W^{Z,N} = 
-2 \left( Z g_{AV}^{ep} + N g_{AV}^{en} \right) \, .
\label{eq:qw-lo}
\end{equation}

There are corrections due to the fact that the matrix elements of the 
electromagnetic and axial-vector neutral current operators between electronic 
states involved in the atomic transition carry a spatial dependence on the 
electric and weak charge distributions in the nucleus \cite{Pollock:1992mv}. 
The dominating part of this correction will be taken into account in the 
calculation by using a non-pointlike charge distribution inside the 
nucleus and assuming that the proton and the neutron distributions are 
the same. However, we will have to add a small correction due to the 
difference between proton and neutron distributions. It turns out that 
this correction is dominated by the difference of radii of the proton 
and neutron distributions, i.e.\ the neutron skin (NSKIN) effect 
\cite{Sil:2005tg}. Therefore, it can be treated as an additive 
contribution  
\begin{equation}
\Delta Q_W^{\rm NS} = 
2 \left(1 - \frac{q_n}{q_p} \right) N g_{AV}^{en} , 
\label{eq:qw-nskin}
\end{equation}
where $q_p$ and $q_n$ are determined from the electronic axial form factor 
weighted by the proton and neutron distributions.  

In the SM, the effective low-energy couplings are determined by the weak neutral-current gauge couplings of the $Z$-boson to quarks and leptons. 
They are fixed by the charge and isospin quantum numbers and the weak mixing angle, $\sin ^{2} \theta_{W}$. We have $g_{A V}^{e u} = - \frac{1}{2} 
+ \frac{4}{3} \sin ^{2} \theta_{W}$, $g_{A V}^{e d} = \frac{1}{2} - \frac{2}{3} \sin ^{2} \theta_{W}$, $g_{V A}^{e u} = - g_{V A}^{e d} 
= \frac{1}{2} - 2 \sin ^{2} \theta_{W}$ and therefore, $g_{AV}^{ep} = - \frac{1}{2} + 2 \sin^2\theta_W $ and $g_{AV}^{en} = \frac{1}{2}$.

At leading-order, the SM predicts relations between the weak mixing angle, the normalization of the effective 4-fermion operators given 
by the Fermi constant and the weak boson masses. The Fermi constant is given by $G_{F} = \pi \alpha / (\sqrt{2} m_{W}^{2} \sin^2\theta_W)$ 
with the fine structure constant $\alpha = e^2 / (4 \pi)$ and the weak mixing angle is related to the gauge boson masses by $m_W = m_{Z} \cos 
\theta_W$. High-precision measurements require to take into account higher-order electroweak radiative corrections, both in these parameter relations 
as well as in predictions for observables. It is convenient to choose the $\overline{\text{MS}}$ renormalization scheme where the weak mixing angle 
becomes a scale-dependent running coupling, usually denoted by $\sin^2\bar{\theta}_W(\mu)$. Effects due to Feynman diagrams with loops ($\gamma Z$ 
mixing, vertex corrections, box graphs) contributing to the observable have been calculated in \cite{Marciano:1982mm} and can be absorbed into 
corrected effective couplings, as described in Ref.~\cite{Erler:2013xha}. Numerically, these corrections can be taken into account by replacing 
Eq.~(\ref{eq:qw-lo}) with 
\begin{eqnarray}
\label{eq:qw-ewnlo}
Q_W^{Z,N} &=& -2 \left(1 - \frac{\alpha}{2\pi} \right) 
\left ( Z \bar{g}_{AV}^{ep} + N \bar{g}_{AV}^{en} \right ) \, , 
\end{eqnarray}
where $\bar{g}_{AV}^{ep} = \rho_{PV} \left(-\frac{1}{2} + 2 \sin^2\bar{\theta}_W(\mu) - 0.00261 \right) - 0.01014$
and $\bar{g}_{AV}^{ep} =  \rho_{PV} \left(\frac{1}{2} - 0.00282 \right) - 0.00242$ with $\rho_{PV} = 1.00063$. The scale $\mu$ has to be chosen 
equal to the typical momentum scale of the experiment. We follow Ref.~\cite{Bouchiat:1983uf} and set $\mu = 2.4$~MeV for Cs, but the precise value 
is not important since $\sin^2\bar{\theta}_W(\mu)$ depends only very weakly on $\mu$ at low scales. 

In a specific model, the coefficients of 4-fermion operators are predictions. They would be related to parameters of an underlying theory. For 
example, 4-fermion contact interactions can originate from Feynman diagrams describing the exchange of a heavy particle at tree-level. Models with 
extra heavy $Z^\prime$-bosons are well-motivated, for example in string-inspired grand unified models with an E$_6$ gauge 
group \cite{London:1986dk}. Spontaneous symmetry breaking generates two extra U(1) factors whose $Z^\prime$-bosons mix 
with each other in general. The lighter of them, with mass $M_{Z_\theta}$, contributes to the weak charge of the nucleon
\begin{equation}
\Delta Q_W^{Z,N}(Z_\theta) = 32 (Z+2N) B (B-A) z ,
\end{equation}
where $z$, $A$ and $B$ are parameters which depend on the mixing angle of the two extra $Z$-bosons, their gauge coupling 
constant $g_\theta$ and on the mass \cite{London:1986dk}. The extra $Z^\prime$ is denoted $Z_\chi$ for the special 
case of no mixing, and one finds \cite{London:1986dk,Marciano:1990dp}
\begin{eqnarray}
\Delta Q_W^{Z,N}(Z_\chi) &=& (Z+2N) \frac{4 \sin^2 \theta_W g_\chi^2}{5 e^2} \frac{M_W^2}{M_{Z_\chi}^2} \nonumber \\
&\simeq & 0.4 (Z+2N) \frac{M_W^2}{M_{Z_\chi}^2} .
\label{massz}
\end{eqnarray}
APV measurements can therefore set a limit on the mass of such an extra heavy $Z^\prime$ boson. 

A different type of heavy new particles without direct couplings to the ordinary fermions can enter at the loop level through 
the $W$ and $Z$ self energies. Examples would be SUSY models at high mass scales, or technicolor models. Just three 
parameters are needed to describe the corresponding effects on observables, usually called $S$, $T$ and $U$ \cite{Peskin:1991sw}. 
We use the definition described in \cite{PDG2020}. Such type of BSM physics can be absorbed in a modification of the 
neutral-current amplitudes by the factor $\rho^{\rm new} = 1 + 0.00782 T$ and by replacing the weak mixing angle 
with $\sin^2 \theta_W \times (1 + 0.0157 S - 0.0112 T)$, where the numerical coefficients in these relations are 
evaluated with the present world-average values of the weak mixing angle and the $W$-boson mass. This results in \cite{Marciano:1990dp} 
\begin{equation} 
\Delta Q_W^{Z,N}(STU) = Z (-0.0145 S + 0.011 T) - N (0.00782 T) . 
\end{equation} 
For ${}^{133}$Cs, this gives 
\begin{equation} 
\Delta Q_W^{55,78}(STU) = Q_W^{55,78} \times (0.0109 S + 0.7 \cdot 10^{-4} T) \, . 
\label{eqs}
\end{equation} 
Thus, APV is sensitive to $S$, the isospin-conserving parameter, while the dependence on the isospin-violating parameter $T$ is very small. 

Finally, we discuss the case of BSM physics at low mass scale which has caused considerable interest recently 
and is motivated by the search for a dark matter particle. A light vector boson associated with a U(1) gauge symmetry 
in the dark sector could couple to ordinary SM matter via kinetic mixing with the photon and mass mixing with the 
SM $Z$-boson \cite{Davoudiasl:2014kua,Davoudiasl:2015bua}. Such a new boson is known as dark-$Z$ boson and could be 
an additional source of parity violation. Its effect can be described by a modification of the running of the weak 
mixing angle in the intermediate to low mass range without visible effects at high-energy $Z$-pole measurements. 
A possible realization of such a scenario can be found with a 2-Higgs doublet model where mixing is generated through 
loop diagrams. The effective weak mixing angle seen at the energy scale $\mu$ would be shifted by \cite{Davoudiasl:2015bua}
\begin{equation}
\Delta \sin^2 \theta_W(\mu) = - \epsilon \delta \frac{M_Z}{M_{Z_d}} \frac{\sin\theta_W \cos\theta_W }{1 + \mu^2 / M^2_{Z_d}} , 
\end{equation}
where $M_{Z_d}$ is the mass of the dark-$Z$, and $\epsilon$ and $\delta$ are model parameters, depending for example 
on the charged Higgs-boson mass if the model is realized with two Higgs doublets. For APV we can assume that $\mu \ll 
M_{Z_d}$ which leaves us with 
\begin{equation}
\Delta \sin^2 \theta_W(\mu) \simeq - 0.43 \epsilon \delta \frac{M_Z}{M_{Z_d}}  . 
\label{eqsnt}
\end{equation}
Several experiments have narrowed down the parameter space for a dark-$Z$ recently, but there is still room for a 
significant modification of $\sin^2 \theta_W$ that can be tested with APV. 

\begin{table}[t]
\caption{Comparison of the calculated energies (in cm$^{-1}$ and $A_{hyf}$ values (in MHz) from the present work with the NIST data and experimental 
results. Since the uncertainties of the experimental (Expt) results are below the significant digits, they are not quoted here.}
\begin{ruledtabular}
\begin{tabular}{lccccc} 
  Method  & $6S$  & $6P_{1/2}$ & $7S$ & $7P_{1/2}$ & $8P_{1/2}$ \\   
 \hline \\ 
 \multicolumn{6}{c}{\underline{Energy values}} \\
This work     & 31357(50) & 20243(20)  & 12861(15)  & 9641(10)  &  5697(10)  \\
Expt \cite{nist} &  31406.47 & 20229.21 & 12871.94 & 9642.12 & 5698.63 \\
 & & & & & \\
 \multicolumn{6}{c}{\underline{$A_{hyf}$ values}} \\
This work  & 2306(10) & 291(2)   & 547(2)   & 94(1)   &  42(1)  \\
Expt & 2298.16$^a$  & 291.91$^b$  & 545.82$^c$ & 94.40$^d$ &  42.97$^e$ \\
\end{tabular}
\end{ruledtabular}
Refs. $^a$\cite{arimondo}; $^b$\cite{dipankar}; $^c$\cite{yang}; $^d$\cite{williams}; $^e$\cite{happer}.\\
\label{tab1}
\end{table}

The NSI neutral current weak interaction Hamiltonian in an atom is given by \cite{Bouchiat}
\begin{eqnarray}
\label{eq1}
H_{APV}^{NSI} &=& 
-\frac{G_F}{2\sqrt{2}}Q_{W} \sum_e \gamma_e^{5} \rho_{nuc}(r_e) ,
\end{eqnarray}
where $\rho_{nuc}(r_e)$ is the electron density within the nucleus 
The charge in the nucleus is described by a Fermi distribution. 
The atomic wave function ($|\Psi_v \rangle$) of a state in the Cs atom 
is calculated by splitting the total Hamiltonian into two parts, 
\begin{eqnarray}
 H = H_{em} + \lambda H_{w}, 
\label{eqham}
\end{eqnarray}
where $H_{em}$ represents the dominant electromagnetic interactions in an atom and $H_{APV}^{NSI} \equiv \lambda H_{w}$ with 
$\lambda=\frac{G_F}{2\sqrt{2}}Q_W^{Z,N}$. We have considered the Dirac-Coulomb-Breit interaction Hamiltonian along with lower-order QED 
corrections due to the self-energy and vacuum polarization effects as $H_{em}$ in our calculations (for details, see \cite{bks-arxiv}). 
Since the strength of $H_{APV}^{NSI}$ is much weaker than that of $H_{em}$ in an atomic system, the wave function $|\Psi_v \rangle$ 
represents a state corresponding to the total Hamiltonian $H=H_{em}+\lambda H_{w}$ and its energy (say, $E_v$) can be expressed as
\begin{eqnarray}
|\Psi_v \rangle \simeq |\Psi_v^{(0)} \rangle + \lambda |\Psi_v^{(1)} \rangle \ \ \text{and} \ \ E_v \simeq E_v^{(0)} + \lambda E_v^{(1)}, \ \ \ \
\end{eqnarray}
where the superscripts 0 and 1 stand for the zeroth-order and first-order contributions due to $H_{w}$, respectively. 
The electric dipole transition amplitude corresponding to two same nominal parity states $|\Psi_i\rangle$ and $|\Psi_f\rangle$ in the presence 
of $H_{APV}^{NSI}$ can be written as \cite{Bouchiat,bouchiat1}
\begin{eqnarray}
 E1_{PV} \simeq \lambda \frac{\langle \Psi_f^{(1)} | D | \Psi_i^{(0)} \rangle + \langle \Psi_f^{(0)} | D | \Psi_i^{(1)} \rangle}{\sqrt{\langle \Psi_f^{(0)} | \Psi_f^{(0)} \rangle \langle \Psi_i^{(0)} | \Psi_i^{(0)} \rangle}}.
\label{e1pnc}
\end{eqnarray}
In the sum-over-states approach, the first-order wave function is expanded as
$|\Psi_v^{(1)} \rangle = \sum_{I \ne v} |\Psi_I^{(0)} \rangle \frac{\langle \Psi_I^{(0)} | H_{w} | \Psi_v^{(0)} \rangle }{E_v^{(0)} - E_I^{(0)}}$,
where $I$ denotes all possible intermediate states, that can be divided into core states (contributions from these states are designated as ``Core''),
low-lying bound states (contributions from these states are given as ``Main''), and the remaining high-lying states including continuum (whose 
contributions are mentioned as ``Tail'') for computational simplicity. 

\begin{table}[t]
\caption{Matrix elements of the operators E1 (in a.u.) and $H_{APV}^{NSI}$ (in units of $-i (Q_{W}/N) \times 10^{-11}$), 
respectively, from our calculations. We also list the precise E1 values inferred from various measurements of lifetimes 
and Stark shifts of atomic states.}
\begin{ruledtabular}
\begin{tabular}{l clc} 
 Transition &  \multicolumn{2}{c}{E1 amplitude} &  $H_{APV}^{NSI}$ amplitude  \\
 \cline{2-3} \\
      &  This work & Experiment & This work \\
 \hline \\
$6P_{1/2} \leftrightarrow 6S$ & 4.5067(40) & 4.5097(74)\cite{Young} & 1.2648(15)  \\
                              &            & 4.4890(65) \cite{Rafac} &   \\
                              &            & 4.505(2) \cite{Patternson}  & \\
                              &            & 4.508(4) \cite{gregoire} & \\
$7P_{1/2} \leftrightarrow 6S$ & 0.2805(20) & 0.2825(20) \cite{Vasilyev} & 0.7210(15)  \\
                              &            &  0.2789(16) \cite{Antypas} & \\
                              &            &  0.27810(45) \cite{damitz} & \\
$8P_{1/2} \leftrightarrow 6S$ & 0.0824(10) &   &  0.4783(10)  \\
$6P_{1/2} \leftrightarrow 7S$ & 4.2559(30) &  4.233(22) \cite{Bouch} & 0.6161(15) \\
                              &            &  4.249(4)  \cite{Toh} &  \\
$7P_{1/2} \leftrightarrow 7S$ & 10.2915(100) & 10.308(15) \cite{Bennett1} & 0.3464(10) \\
$8P_{1/2} \leftrightarrow 7S$ & 0.9623(20)  &   & 0.2296(05) \\
 \end{tabular}
\end{ruledtabular}
\label{tab2}
\end{table}

\begin{table*}[t]
\caption{The `Core', `Main' and `Tail' contributions to the $E1_{PV}$ amplitude (in 
units of $-i (Q_{W}/N) ea_0 \times 10^{-11}$) using the Dirac-Coulomb 
Hamiltonian in the DHF, RCCSD and RCCSDT methods. The `Main' contribution is determined using the $np ~ ^2P_{1/2}$ intermediate 
states with $n=6$, 7 and 8. Contributions from Breit and QED interactions are quoted separately. Contributions from ``Extra'', the neutral weak 
interactions among electrons ($e-e$) and the NSKIN effect are also mentioned. The final results (Final) from different works show significant differences.}
 \begin{ruledtabular}
  \begin{tabular}{l c c c c c c  c cc}
  Method    & Core & Main & Tail &  Breit  &  QED &  Extra & $e-e$ & $\delta E1_{PV}^{NS}$ & Final\\
 \hline \\
 DHF & $-0.0017$  & 0.7264 & 0.0137  &  &  &  & \\ 
 RCCSD & $-0.0019$ & 0.8623 & 0.0357 &  &  &  &  & \\
 RCCSDT & $-0.0018$ & 0.8594 & 0.0391$^a$ & $-0.0055$ & $-0.0028$ & 0.0026 & 0.0003$^b$ & $-0.00377(39)$ & 0.8893(27) \\
 \hline \\ 
 Ref. \cite{dzuba1} & $0.0018(8)$ & 0.8823(17)$^{a,b}$ & 0.0238(35) & $-0.0055(1)^b$ & $-0.0029(3)^b$ & &  & $-0.0018(5)^{b}$ &  0.8977(40) \\
 Ref. \cite{porsev1} & $-0.0020$ & 0.8823(17)$^a$ & 0.0195 & $-0.0054^b$  & $-0.0024^b$  &  $-0.00006$  & 0.0003$^b$ & $-0.0017^{b}$ & 0.8906(24) \\
 Ref. \cite{blundell} & $-0.002(2)$ & 0.893(7)$^a$  &  0.018(5) & $-0.002(2)$ &  &  &  & $-0.0006$ & 0.907(9) \\
  \end{tabular}
 \end{ruledtabular}
 \label{tab3}
 $^a$ Contains additional contribution from the $9p ~ ^2P_{1/2}$ state. $^b$Taken from previous calculation \cite{Milstein}.
\end{table*}

The latest two high precision calculations, reported in Refs.~\cite{porsev1,dzuba1}, are carried out by estimating the ``Core'', ``Main'' and ``Tail''
contributions by applying mixed many-body methods. The calculations in Ref. \cite{porsev1} included the valence triple excitation effects to 
``Main'' by employing the relativistic coupled-cluster (RCC) theory, and it was found that these effects to the atomic properties of $^{133}$Cs 
were relatively important in reducing the uncertainty in the $E1_{PV}$ amplitude to 0.27\% \cite{porsev1}. This result was in good agreement with the
SM, however the calculation on which it is based had used a sum-over-states approach in which the `Main' contributions were estimated only from the 
excited states up to the principal quantum number $n=9$. Later Dzuba {\it et al.} reported another result in Ref.~\cite{dzuba1} with 0.5\% accuracy 
by using the ``Main'' contribution from Ref.~\cite{porsev1}, but with different ``Core'' (opposite sign than \cite{porsev1}) and ``Tail'' 
contributions by taking into account certain sub-classes of correlation effects. They found substantial differences in these contributions from 
Ref. \cite{porsev1}; especially the ``Core'' contribution differed by about 200\% (due to opposite sign). This resulted in 0.8\% difference between 
the final results of Porsev {\it et al}.~\cite{porsev1} and Dzuba 
{\it et al}.~\cite{dzuba1}. In addition, both the above works did not include double
core-polarization (DCP) effects \cite{roberts2}, and contributions from the Breit and QED effects were taken from the earlier works. To include all 
these neglected contributions and to treat all the electron correlation effects on an equal footing, we solve the inhomogeneous equation for the 
first-order wave function
\begin{equation}
 (H_{em}-E_v^{(0)}) |\Psi_v^{(1)} \rangle = (E_v^{(1)}-H_{w})|\Psi_v^{(0)} \rangle ,
 \label{eqpt}
\end{equation}
where $E_v^{(1)}=0$ in the present case owing to the odd-parity nature of $H_{w}$. This is achieved by expressing the unperturbed and the first-order 
wave function of the Cs atom in the RCC theory framework as \cite{bijaya1,bijaya2,bijaya3}
\begin{eqnarray}
 |\Psi_v^{(0)} \rangle &=& e^{T^{(0)}} \left \{ 1+ S_v^{(0)} \right \} |\Phi_v \rangle , \\ 
\text{and} \ \ |\Psi_v^{(1)} \rangle &=& e^{T^{(0)}} \left \{ S_v^{(1)}+ T^{(1)} \left (1+S_v^{(0)} \right ) \right \} |\Phi_v \rangle, \label{eqcc} \ \ \ \ \
\end{eqnarray}
where $|\Phi_v \rangle$ is obtained by determining the Dirac-Hartree-Fock (DHF) wave function of the closed-core ($|\Phi_0 \rangle$) and then, appending the 
corresponding valence orbital $v$ to it as $|\Phi_v \rangle= a_v^{\dagger}|\Phi_0 \rangle$. $T^{(0)}$ and $S_v^{(0)}$ are the core and the valence 
excitation operators. The superscript 0 represents the absence of any external perturbation. Similarly, $T^{(1)}$ and $S_v^{(1)}$ are the core and 
the valence excitation operators with the superscript 1 representing the order of perturbation in $H_{w}$. In our previous calculations, we had successfully
employed this approach based on RCC theory with singles and doubles approximation (RCCSD method) for the evaluation of $E1_{PV}$ amplitudes in Ba$^+$ 
\cite{bijaya1}, Ra$^+$ \cite{bijaya2} and Yb$^+$ \cite{bijaya3} and had achieved results within 1\% accuracy. In the present work, we have implemented 
additional triple excitations beyond the RCCSD method (RCCSDT method) to achieve sub-one percent accurate $E1_{PV}$ in $^{133}$Cs as there is a 
renewed interest in the inclusion of the neglected correlation effects in this atom (e.g. see discussions in \cite{derevianko1,derevianko2}). It is 
worth mentioning here that we excite all the electrons in the RCCSD method to account for the electron correlation effects, but correlate all the 
electrons except from the $1-3s$, $2-3p$, and $3d$ occupied orbitals and beyond $n=15$ virtual orbitals for triple excitations due to limitations in 
the available computational resources. Also, we have considered active orbitals up to $l=5$ in our RCC calculations and contributions from the 
orbitals belonging to higher angular momentum symmetries, quoted as `Extra' hereafter, are estimated using low-order perturbative methods. 

We first calculated the energies, electric dipole (E1) matrix elements and magnetic dipole hyperfine structure constants ($A_{hyf}$)
of the states that give rise to dominant contributions to the determination of $E1_{PV}$ in $^{133}$Cs. By comparing these values with their 
corresponding experimental results \cite{nist,arimondo,dipankar,yang,williams,happer,Young, Rafac,Patternson,gregoire,Vasilyev,Antypas,damitz,Bouch,
Toh,Bennett1}, we have assessed the accuracies of the wave functions in the regions close to and far away from the nucleus. Our calculated 
values of these properties at different levels of approximations can be found in Ref.~\cite{bks-arxiv}, however, the final values along with their 
experimental results are listed in Tables \ref{tab1} and \ref{tab2}. As can be seen from these two tables, comparison of our calculations of the above
properties with their precisely known experimental data is very impressive and within sub-one percent accuracy. This strongly suggests that our 
atomic wave functions are very reliable and they can be used to determine the transition amplitude $E1_{PV}$  accurately. 

Keeping in mind our classification of the RCC terms, we find  the `Core'  contribution to $E1_{PV}$ and also the `Main' contribution using our calculated 
energies, E1 matrix elements and amplitudes of $H_{APV}^{NSI}$ for the intermediate $n(=6,7,8)P_{1/2}$ states that are quoted in Tables \ref{tab1} and 
\ref{tab2}. After subtracting the ``Core'' and ``Main'' contributions from the final value, the remainder is taken as
the ``Tail'' contribution. These contributions from the DHF, RCCSD and RCCSDT methods using the Dirac-Coulomb Hamiltonian are quoted in Table \ref{tab3}. 
In addition, we give contributions from the Breit and QED interactions of our calculation using the RCCSDT method in the same table. The other 
neglected contributions due to `Extra', possible neutral weak interactions among electrons ($e-e$), and the NSKIN effect that 
were not included in our RCC calculation, are also quoted in the above table. The small $e-e$ contribution to $E1_{PV}$ has been taken from Ref. 
\cite{Milstein}. 

In a seminal work, Fortson et al \cite{fortson} had analyzed the NSKIN effect on APV. By adopting their analysis, 
the effect of NSKIN on $E1_{PV}$ ($\delta E1_{PV}^{NS}$) can be estimated by
\begin{eqnarray}
 \delta E1_{PV}^{NS}(~^{133}\text{Cs}) \approx - \frac{3}{7} (\alpha Z)^2 \ t \ E1_{PV} ,
\end{eqnarray}
where $t$ is known as the neutron skin parameter which describes 
the relative difference of the r.m.s.\ radii of the neutron and 
proton distributions in the nucleus. This empirical formula was used in Refs. \cite{porsev1,dzuba1} for determining $\delta E1_{PV}^{NS}$. 
Using $t=0.033(8)$ \cite{brown} and the uncorrected $E1_{PV}$ value 0.8914, we get $\delta E1_{PV}^{NS} \simeq -0.0020(5)$ in units of 
$\times 10^{-11} i (-Q_{W}/N)ea_0$. This is in good agreement with the recently estimated value by Brown et al. in Ref.~\cite{brown}. 
However, Sil et al.\ have estimated $\Delta Q_W^{\rm NS}$ by employing a more rigorous effective field theory framework \cite{Sil:2005tg}. 
We use the relation 
\begin{eqnarray}
 \delta E1_{PV}^{NS}(~^{133}\text{Cs}) 
 &\approx & \frac{\Delta Q_W^{\rm NS}}{Q_W} E1_{PV} 
\end{eqnarray}
and the numerical results from Ref.~\cite{Sil:2005tg}. Interpreting the two model results considered there to define a central value and an uncertainty 
range, we find $\delta E1_{PV}^{NS} \simeq-0.00377(39) \times 10^{-11} i (-Q_{W} / N)ea_0$ by substituting $Q_W \simeq -73.23$. This is a slightly 
larger correction than considered in the previous Refs.~\cite{porsev1,dzuba1,blundell}.

We compare individual contributions with the previously reported RCC results using the sum-over-states approach \cite{porsev1,blundell} and the 
latest reported result \cite{dzuba1} of $E1_{PV}$ in Cs. These calculations include the $9P_{1/2}$ state  in their ``Main'' contribution in the 
sum-over-states approach, whereas our ``Tail'' includes the contribution from this state. Our final $E1_{PV}$ value is 0.8893(27) in contrast to 
the results that have been reported previously as 0.8906(24) \cite{porsev1} and 0.8977(40) \cite{dzuba1} in units of $\times 10^{-11} i (-Q_{W}/N) 
ea_0$. The major difference between the results from Ref. \cite{porsev1} and ours is because of the fact that they account for
different NSKIN effect. Large difference between the present calculation and that of Ref. \cite{dzuba1} is mainly due to the `Core' contributions, 
which have different signs in both the cases. 

Nonetheless, one of the most important achievements of our calculation is that it resolves the ambiguity of the sign  of the``Core'' contribution from 
the calculations reported in Refs. \cite{porsev1} and \cite{dzuba1}. We have adopted the same procedure as in Ref. \cite{porsev1} to estimate the 
uncertainty of $E1_{PV}$. This is also independently verified by analyzing major sources of uncertainties that can come from the neglected higher level
excitations in the RCC theory and finite-size basis functions used in the calculation. If the difference between the RCCSD and RCCSDT values is assumed 
to be the maximum contribution from the neglected higher level excitations and considering `Extra' as the maximum uncertainty due to incompleteness in 
the used basis functions, we also arrive at the same uncertainty of $E1_{PV}$.

It is necessary to combine our $E1_{PV}$ value with the precisely measured $Im(E1_{PV}/\beta)=1.5935(56)$ mV/cm \cite{wood}, where $Im$ refers to the
imaginary part and $\beta$ is the vector polarizability of the $6s ~ ^2S_{1/2} - 7s ~ ^2S_{1/2}$ transition in $^{133}$Cs to extract $Q_W^{Z,N}$. 
However, this also requires an accurate knowledge of $\beta$. We have determined this quantity by using the precisely available E1 matrix elements 
either from measurements or from our calculation as discussed in detail in Ref. \cite{bks-arxiv} and obtain its value as $\beta=27.12(4) \ ea_0^3$. 
Using all these values, we get $Q_W^{Z,N}=-73.71(26)_{ex}(23)_{th}$. This is in agreement with the SM prediction $Q_W^{\text{SM}} = 
-73.23(1)$, obtained from Eq.~(\ref{eq:qw-ewnlo}) with $\sin^2\bar{\theta}_W(2.4\,\text{MeV}) = 0.23857(5)$ \cite{PDG2020}. 
At the 1$\sigma$ confidence level, we have $\Delta Q_W^{Z,N} \equiv Q_W^{Z,N} - Q_W^{\text{SM}}=-0.48(35)$. In turn, our value of $Q_W^{Z,N}$ can be 
used to determine the weak mixing angle. We find $\sin^2\bar{\theta}_W(2.4\,\text{MeV}) = 0.2408(16)$ with a slightly smaller uncertainty and a 
significant shift of the central value compared with the previous determination \cite{PDG2020}. 

The experimental value of $Q_W^{Z,N}$ provides a constraint on the low-energy effective electron-quark couplings: $376 g_{AV}^{eu} + 422 g_{AV}^{ed} 
= 73.71(35)$. Assuming the SM prediction for one of them, we find a value for the other: $g_{AV}^{eu} = -0.1877(9)$ for $g_{AV}^{ed} = 0.3419$ and 
$g_{AV}^{ed} = 0.3429(8)$ for $g_{AV}^{eu} = -0.1888$. We also find slightly improved limits on the BSM parameters described in the introduction. 
The isospin conserving oblique parameter $S$ can be constrained to $S \simeq 0.60(44)$ assuming $T=0$ in Eq.~(\ref{eqs}). The central value of $S$ is 
shifted to positive values, compared with the previous determination $S \simeq -0.51(52)$ \cite{PDG2020}. From Eq.~(\ref{massz}), we obtain a limit on the
mass of an extra $Z^\prime$ boson. Since the shift is always positive, we find a one-sided exclusion limit at 95~\% confidence limit of $M_{Z_x} > 2.36$~TeV, using $M_W = 80.379$ GeV \cite{PDG2020}, 
compared with recent limits from the ATLAS collaboration who found values ranging from 3.5 to 4.5~TeV \cite{Aaboud:2018jff}. Furthermore, using 
Eq.~(\ref{eqsnt}) we can constrain the dark-$Z$ model parameter $\epsilon \delta \frac{M_Z}{M_{Z_d}} \simeq - 0.0051(37)$. 

In conclusion, we used a perturbed version of the relativistic coupled-cluster theory to calculate the nuclear spin-independent parity violating 
electric dipole transition amplitude for the $6s ~ ^2S_{1/2} - 7s ~ ^2S_{1/2}$ transition in $^{133}$Cs. The principal merit of this approach is that
it treats the contributions of the core, valence and excited states to the above parity violating transition amplitude on the same footing, thereby
overcoming the limitations of the previous high precision calculations of this quantity. Our work resolved the ambiguity in the sign difference for 
the contribution from core states. In addition, we estimated the uncertainty in our calculation of the parity violating transition amplitude in
$^{133}$Cs more rigorously than those in previous calculations. The salient implications of the deviation of the nuclear weak charge from the standard
model, that is obtained in the present work, for probing possible new physics have been discussed. Our result, in combination with 
measurements from proposed new high-precision experiments, has the potential to improve the constraints on beyond the standard 
model physics in the future.

We acknowledge support by the Mainz Institute for Theoretical Physics (MITP). This work was initiated during the MITP Virtual 
Workshop ``Parity Violation and Related Topics''. Computations reported in this work were performed using the PRL Vikram-100 HPC cluster.


\begin{thebibliography}{99}
 
\bibitem{Bouchiat}
M.-A. Bouchiat and C. Bouchiat, Phys. Lett. B {\bf 48}, 111 (1974); J. Phys. (France) {\bf 35}, 899 (1974).

\bibitem{wood}
C. S. Wood, S. C. Bennet, D. Cho, B. P. Masterson, J. L. Roberts, C. E. Tanner, and C. E. Wieman, Science {\bf 275}, 1759 (1997).

\bibitem{choi}
J. Choi and D. S. Elliott, Phys. Rev. A {\bf 93}, 023432 (2016).

\bibitem{anders}
A. Kastberg, T. Aoki, B. K. Sahoo, Y. Sakemi and B. P. Das, Phys. Rev. A {\bf 100}, 050101(R) (2019).

\bibitem{bouchiat1}
M.-A. Bouchiat and C. Bouchiat, Rep. Prog. Phys. {\bf 60}, 1351 (1997).

\bibitem{PDG2020} 
P. A. Zyla et al. [Particle Data Group], Prog. Theor. Exp. Phys. {\bf 2020}, 083C01 (2020). 

\bibitem{Beringer:1900zz}
J. Beringer \textit{et al.} [Particle Data Group], Phys. Rev. D \textbf{86}, 010001 (2012).

\bibitem{Pollock:1992mv}
S. J. Pollock, E. N. Fortson and L. Wilets, Phys. Rev. C \textbf{46}, 2587 (1992).

\bibitem{Sil:2005tg}
T. Sil, M. Centelles, X. Vinas and J. Piekarewicz, Phys. Rev. C \textbf{71}, 045502 (2005).

\bibitem{Marciano:1982mm}
W.~J.~Marciano and A.~Sirlin, Phys. Rev. D \textbf{27}, 552 (1983). 

\bibitem{Erler:2013xha}
J. Erler and S. Su, Prog. Part. Nucl. Phys. \textbf{71}, 119 (2013).

\bibitem{Bouchiat:1983uf}
C. Bouchiat and C. A. Piketty, Phys. Lett. B \textbf{128}, 73 (1983). 

\bibitem{London:1986dk}
D. London and J. L. Rosner, Phys. Rev. D \textbf{34}, 1530 (1986). 

\bibitem{Marciano:1990dp}
W. J. Marciano and J. L. Rosner, Phys. Rev. Lett. \textbf{65}, 2963 (1990); [erratum: Phys. Rev. Lett. \textbf{68}, 898 (1992)]. 

\bibitem{Peskin:1991sw}
M. E. Peskin and T. Takeuchi, Phys. Rev. D \textbf{46}, 381 (1992). 

\bibitem{Davoudiasl:2014kua}
H. Davoudiasl, H. S. Lee and W. J. Marciano, Phys. Rev. D \textbf{89}, 095006 (2014).

\bibitem{Davoudiasl:2015bua}
H. Davoudiasl, H. S. Lee and W. J. Marciano, Phys. Rev. D \textbf{92}, 055005 (2015).

\bibitem{bks-arxiv}
B. K. Sahoo and B. P. Das, arXiv:2008.0891 (unpublished).
  
\bibitem{porsev1}
S. G. Porsev, K. Beloy, and A. Derevianko, Phys. Rev. Lett. {\bf 102}, 181601 (2009); Phys. Rev. D {\bf 82}, 036008 (2010).
  
\bibitem{dzuba1}
V. A. Dzuba, J. C. Berengut, V. V. Flambaum, and B. Roberts, Phys. Rev. Lett. {\bf 109}, 203003 (2012).
  
\bibitem{roberts2}
B. M. Roberts, V. A. Dzuba and V. V. Flambaum, Phys. Rev. A {\bf 88}, 042507 (2013).

\bibitem{bijaya1}
B. K. Sahoo, R. K. Chaudhuri, B. P. Das and D. Mukherjee, Phys. Rev. Lett. {\bf 96}, 163003 (2006).

\bibitem{bijaya2}
L. Wansbeek, B. K. Sahoo, R. G. E. Timmermans, K. Jungmann, B. P. Das, and D. Mukherjee, Phys. Rev. A {\bf 78}, 050501(R) (2008).

\bibitem{bijaya3}
B. K. Sahoo and B. P. Das, Phys. Rev. A {\bf 84}, 010502(R) (2011).  

\bibitem{derevianko1}
M. S. Safronova, D. Budker, D. DeMille, D. F. J. Kimball, A. Derevianko and C. W. Clark, Rev. of Mod. Phys. {\bf 90}, 025008 (2018).

\bibitem{derevianko2}
C. Wieman and A. Derevianko, arXiv:1904.00281

\bibitem{nist}
A. Kramida, Yu. Ralchenko, J. Reader, and NIST ASD Team (2018), {\it NIST Atomic Spectra Database} (ver. 5.6.1), 
 National Institute of Standards and Technology, Gaithersburg, MD.

\bibitem{arimondo}
E. Arimondo, M. Inguscio, and P. Violino, Rev. Mod. Phys. {\bf 49}, 31 (1977).

\bibitem{dipankar}
D. Das and V. Natarajan, J. Phys. B {\bf 39}, 2013 (2006).

\bibitem{yang}
Guang Yang, Jie Wang, Baodong Yang, and Junmin Wang, Laser Phys. Letts. {\bf 13}, 085702 (2016).

\bibitem{williams}
W. D. Williams, M. T. Herd and W. B. Hawkins, Laser Phys. Letts. {\bf 15}, 095702 (2018).

\bibitem{happer}
W. Happer, {\it Atomic Physics 4}, eds. G. zu Putlitz, E. W. Weber, and A. Winnacker, (Plenum Press, New York) pp. 651-682 (1974).

\bibitem{Young}
L. Young, W. T. Hill, III, S. J. Sibener, S. D. Price, C. E. Tanner, C. E. Wieman and S. R. Leone, Phys. Rev. A {\bf 50}, 2174 (1994).

\bibitem{Rafac}
R. J. Rafac, C. E. Tanner, A. E. Livingston, and H. G. Berry, Phys. Rev. A {\bf 60}, 3648 (1999).

\bibitem{Patternson}
B. M. Patterson, J. F. Sell, T. Ehrenreich, M. A. Gearba, G. M. Brooke, J. Scoville, and R. J. Knize, Phys. Rev. A {\bf 91}, 012506 (2015).

\bibitem{gregoire}
M. D. Gregoire, I. Hromada, W. F. Holmgren, R. Trubko, and A. D. Cronin, Phys. Rev. A {\bf 92}, 052513 (2015).

\bibitem{Vasilyev}
A. A. Vasilyev, I. M. Savukov, M. S. Safronova, and H. G. Berry, Phys. Rev. A {\bf 66}, 020101 (2002).

\bibitem{Antypas}
D. Antypas and D. S. Elliott, Phys. Rev. {\bf 88}, 052516 (2013).

\bibitem{damitz}
A. Damitz, G. Toh, E. Putney, C. E. Tanner and D. S. Elliott, Phys. Rev. A {\bf 99}, 062510 (2019).

\bibitem{Bouch}
M. -A. Bouchiat, J. Guena, and L. Pottier, J. Phys. (Paris), Lett. {\bf 45}, 523 (1984).

\bibitem{Toh}
G. Toh, A. Damitz, N. Glotzbach, J. Quirk, I. C. Stevenson, J. Choi, M. S. Safronova and D. S. Elliot, Phys. Rev. A {\bf 99}, 032504 (2019).

\bibitem{Bennett1}
S. C. Bennett, J. L. Roberts, and C. E. Wieman, Phys. Rev. A {\bf 59}, R16 (1999).

\bibitem{Milstein}
A. I. Milstein, O. P. Sushkov, and I. S. Terekhov, Phys. Rev. Lett. {\bf 89}, 283003 (2002).

\bibitem{fortson}
E. N. Fortson, Y. Pang, and L. Wilets, Phys. Rev. Lett. {\bf 65}, 2857 (1990).

\bibitem{brown}
B. A. Brown, A. Derevianko, and V. V. Flambaum, Phys. Rev. C {\bf 79}, 035501 (2009).

\bibitem{blundell}
S. A. Blundell, W. R. Johnson, and J. Sapirstein, Phys. Rev. Lett. {\bf 65}, 1411 (1990); Phys. Rev. D {\bf 45}, 1602 (1992).

\bibitem{Aaboud:2018jff}
M.~Aaboud \textit{et al.} [ATLAS], Phys. Rev. D \textbf{98}, 092008 (2018). 

  
\end{thebibliography}
\end{document}